# Ultrafast Opto-magnetic Effects in the Extreme Ultraviolet Spectral Range

Martin Hennecke,[1,*] Clemens von Korff Schmising,[1] Kelvin Yao,[1] Emmanuelle Jal,[2] Boris Vodungbo,[2] Valentin Chardonnet,[2] Katherine Légaré,[3] Flavio Capotondi,[4] Denys Naumenko,[4] Emanuele Pedersoli,[4] Ignacio Lopez-Quintas,[4,5] Ivaylo P. Nikolov,[4] Lorenzo Raimondi,[4] Giovanni De Ninno,[4,6] Leandro Salemi,[7] Sergiu Ruta,[8,9] Roy Chantrell,[8] Thomas Ostler,[10] Bastian Pfau,[1] Dieter Engel,[1] Peter M. Oppeneer,[7] Stefan Eisebitt,[1,11] Ilie Radu[1,12,*]

[1] Max-Born-Institut für Nichtlineare Optik und Kurzzeitspektroskopie, Max-Born-Straße 2A, 12489 Berlin, Germany.

[2] Sorbonne Université, CNRS, Laboratoire de Chimie Physique -- Matière et Rayonnement, LCPMR, 75005 Paris, France.

[3] Institut National de la Recherche Scientifique, INRS-EMT, Varennes, Québec J3X 1P7, Canada.

[4] FERMI, Elettra-Sincrotrone Trieste, 34149 Basovizza, Trieste, Italy.

[5] Grupo de Investigación en Aplicaciones del Láser y Fotónica, Departamento de Física Aplicada, University of Salamanca, 37008 Salamanca, Spain.

[6] Laboratory of Quantum Optics, University of Nova Gorica, 5001 Nova Gorica, Slovenia.

[7] Department of Physics and Astronomy, Uppsala University, P.O. Box 516, SE-751 20 Uppsala, Sweden.

[8] Department of Physics, University of York, York YO10 5DD, United Kingdom.

[9] College of Business, Technology and Engineering, Sheffield Hallam University, Howard Street, Sheffield S1 1WB, United Kingdom.

[10] Department of Physics and Mathematics, University of Hull, Cottingham Road, Hull HU6 7RX, United Kingdom.

[11] Institut für Optik und Atomare Physik, Technische Universität Berlin, Straße des 17. Juni 135, 10623 Berlin, Germany.

[12] European X-ray Free-Electron Laser, Holzkoppel 4, 22869 Schenefeld, Germany.

* hennecke@mbi-berlin.de; ilie.radu@xfel.eu

**Abstract**

Coherent light-matter interactions mediated by opto-magnetic phenomena like the inverse Faraday effect (IFE) are expected to provide a non-thermal pathway for ultrafast manipulation of magnetism on timescales as short as the excitation pulse itself. As the IFE scales with the spin-orbit coupling strength of the involved electronic states, photo-exciting the strongly spin-orbit coupled core-level electrons in magnetic materials appears as an appealing method to transiently generate large opto-magnetic moments. Here, we investigate this scenario in a ferrimagnetic GdFeCo alloy by using intense and circularly polarized pulses of extreme ultraviolet radiation. Our results reveal ultrafast and strong helicity-dependent magnetic effects which are in line with the characteristic fingerprints of an IFE, corroborated by *ab initio* opto-magnetic IFE theory and atomistic spin dynamics simulations.

**Introduction**

Light-driven ultrafast magnetic phenomena occurring on ever faster and smaller time and length scales are at the core of the modern magnetism research[1,2]. A long sought-after and yet to be realized phenomenon is the coherent and deterministic control of a macroscopically ordered spin ensemble on the sub-cycle timescales of the photo-exciting laser field[3,4]. In this context, one of the most prominent effects that could lead to large, ultrafast changes of magnetization is the inverse Faraday effect (IFE). Discovered originally in paramagnetic solids[5], the IFE was shown to generate an effective magnetic field or, equivalently, an induced magnetic moment in a medium upon photoexcitation with circularly polarized light pulses. According to classical IFE theory[5], the orientation and the magnitude of the light-induced magnetic moment are respectively determined by the helicity of circularly polarized light (or its angular momentum) and the combined effect of the magneto-optical susceptibility of the photo-excited medium and the incident light intensity. In case of metallic ferro- or antiferromagnetic materials, a fully *ab initio* quantum-mechanical treatment of the IFE has revealed that its action on these materials is significantly different compared to the originally discovered effect, predicting a complex dependence of the IFE-induced magnetization on both helicity and wavelength, which is not covered anymore by the conventional theory[6–8]. However, the high absorption of the optical excitation in these materials typically leads to strong heat-induced demagnetization which hinders the observation of any potential non-thermally induced magnetization[9]. Thus, the actual influence of an IFE on the sub-picosecond spin dynamics observed in such systems as, for example, all-optical switching phenomena in ferrimagnetic rare-earth transition-metal alloys like GdFeCo[10–13], remains highly debated and elusive so far[14].

The magnitude of the IFE depends on the light intensity and its polarization, as well as on the spin-orbit coupling (SOC) of the involved electronic states via the wavelength- and helicity-dependent opto-magnetic constant $\mathcal{K}_{\sigma_\pm}^{\text{IFE}}(\omega)$ [6,8]:

$$\vec{M}_{\text{ind}}(\omega) = i\mathcal{K}_{\sigma_\pm}^{\text{IFE}}(\omega)\left[\vec{E}(\omega) \times \vec{E}^*(\omega)\right], \tag{1}$$

where $\vec{E}(\omega)$ denotes the light electric field, $\sigma_\pm$ its helicity, and $\vec{M}_{\text{ind}}$ the induced magnetization along the $\vec{k}$ vector of the incident circularly polarized light beam. The IFE, which is an electronic Raman process, has thus far been investigated theoretically[6,8,15–18], but a direct comparison with measurements is lacking. A potential way of generating strong opto-magnetic interactions in metallic magnets is the excitation of core-level electrons, involving states with much higher SOC compared to the valence band (e.g., 1.1 eV vs. 65 meV in case of Fe 3p and 3d electrons, respectively[19]) that can be accessed by short wavelength radiation in the extreme ultraviolet (XUV) or soft x-ray range. The availability of free-electron lasers (FEL) providing femtosecond pulses of very high brilliance with control over the light polarization has enabled studies of x-ray induced magnetization dynamics, extending the capabilities from element-specific probing, employing, e.g., the x-ray magnetic circular dichroism (XMCD), towards element-specific excitation of highly non-equilibrium states[20–24]. However, besides first theoretical predictions, there is, to the best of our knowledge, no experimental evidence available so far on the existence of opto-magnetic effects like the IFE at these XUV wavelengths.

In this work, we study the ultrafast magnetization dynamics of ferrimagnetic GdFeCo alloy induced by photoexcitation with intense femtosecond circularly polarized pulses in the XUV spectral range generated by the free-electron laser FERMI. Comparing the helicity-dependent dynamics upon on- and off-resonant excitation with respect to the Fe $M_{3,2}$ resonance, we find strong dynamic helicity- and wavelength-dependent effects that resemble the expected characteristics of an IFE. Analyzing the magnitude and wavelength-dependence of the observed effect, we can rule out a thermal origin caused by dichroic absorption due to the XMCD. This finding is corroborated by atomistic spin dynamics (ASD) simulations, revealing that even in case of fully on-resonant excitation, the expected XMCD response is up to an order of magnitude smaller compared to the observed effect. Instead, we find qualitative agreement with *ab initio* calculations of the IFE in the XUV spectral range. Our results thus identify the existence of an IFE in the XUV spectral range.

## Results and discussion

### Time-resolved pump – probe measurements

The experimental concept employing the XUV pump − magneto-optics probe technique to trigger and measure the IFE is depicted in Fig. 1a. The sample was excited at normal incidence using ≈90 fs (full width at half maximum, FWHM) XUV pulses of variable polarization [circular right/left ($\sigma_\pm$) and linear horizontal (lin. hor.)], tuned to photon energies around the Fe $M_{3,2}$ resonance. The XUV-induced dynamics were probed under 45° by linearly polarized ≈90 fs (FWHM) optical pulses at 400 nm wavelength, measuring both the magneto-optical Faraday rotation in transmission and the Kerr rotation in reflection, obtaining magnetic contrast by flipping an out-of-plane saturating magnetic field of $\pm 8$ mT applied to the sample.

Fig. 1b shows the XUV absorption spectra (XAS) measured on the GdFeCo sample across the Fe $M_{3,2}$ resonance for opposite magnetic fields and the resulting XMCD spectrum. The absorption peak at around 56.1 eV corresponds to the resonant excitation of 3p electrons of Fe into 3d states; the second and less pronounced peak around 62.0 eV arises due to resonant excitation of the small fraction of Co atoms. Note that the vanishing XMCD at this photon energy results from the superposition of bipolar XMCD features of the Fe and Co $M_{3,2}$ resonances weighted by their different concentrations in the alloy (compare Supplementary Figure 8). To study the wavelength-dependence of the XUV excitation across the Fe $M_{3,2}$ resonance, the photon energy of the FEL pulses was tuned to fixed positions in the spectrum at 51.0 eV (below resonance), 54.1 eV (largest XMCD), 56.1 eV (highest absorption) and 64.0 eV (above resonance).

Fig. 2 shows the temporal evolution of magnetization after resonant excitation by linearly and circularly polarized XUV pulses at a photon energy of 54.1 eV, as measured by the Faraday signal; for the Kerr measurements leading to similar results, see Supplementary Note 3. The data was fitted using a double-exponential fit function convoluted with a Gaussian (solid lines in Fig. 2), taking into account the ultrafast demagnetization, the subsequent relaxation process and the experimental time resolution of ≈280 fs. For clarity, the pump-probe scans are displayed only for two incident XUV fluences, showing the dynamics upon low and high fluence excitation. The data reveals a clear influence of the XUV polarization on the demagnetization amplitudes, emerging upon excitation with increasing incident fluence (compare 1.3 and 6.7 mJ/cm$^2$). Above 1.3 mJ/cm$^2$, the XUV pump helicity drives either an enhancement or a decrease of the demagnetization effect that takes place within the first hundreds of femtoseconds; this helicity-dependent difference in demagnetization magnitudes lasts for tens of picoseconds, i.e., also during the magnetization relaxation process, emphasizing the strength and robustness of the transient XUV-induced magnetization. Such a helicity-dependent behavior is at variance with the case of linearly polarized XUV excitation that generates demagnetization based on purely thermal effects, i.e., demagnetization driven by ultrafast electronic heating.

**Fluence- and wavelength-dependence**

In Fig. 3, the demagnetization amplitudes obtained from the double-exponential fits are shown as a function of XUV photon energy, polarization, and incident excitation fluence, evaluated for the full range of used XUV fluences. The data reveals a strong helicity-dependent effect that is present over the whole range of photon energies used for excitation, be it on- (54.1 and 56.1 eV) or off-resonant (51.0 and 64.0 eV) with respect to the Fe $M_{3,2}$ resonance. The splitting between either enhanced or attenuated demagnetization amplitudes upon $\sigma_\pm$-polarized excitation ($D = 1-\min[M/M_0]$, Fig. 3a) and accordingly the magnitude and sign of the difference ($\Delta M = D[\sigma_-]-D[\sigma_+]$, Fig. 3b) show that the effect is wavelength-dependent, clearly changing its polarity along the resonance. The data further reveals that the effect is strongly fluence-dependent; within the experimental uncertainty, two fluence regimes become apparent (illustrated by the white/grey shaded background in Fig. 3b) where the magnitude of $\Delta M$ initially increases with XUV fluence and then either reaches a saturated state at a constant $\Delta M$ value or decreases again as the fluence is further elevated. In case of below- and above-resonant excitation (51.0 and 64.0 eV), $\Delta M$ scales linearly with the fluence until the saturated state is achieved. It thereby reaches up to (32.2±0.7) % difference in pump-induced change with respect to the equilibrium magnetization (largest $\Delta M$ observed for 5.3 mJ/cm$^2$ at 64.0 eV). In case of resonant excitation (54.1 and 56.1 eV), the data reveals a more complex fluence-dependence, indicating either a threshold-like behavior (54.1 eV) or even a fluence-dependent sign change of $\Delta M$ (56.1 eV). For excitation between 54.1 and 64.0 eV, the demagnetization reaches amplitudes of up to ≈100% at high fluences (depending on the helicity), corresponding to full demagnetization of the magnetic layer. This leads to the observed saturation or even quenching of $\Delta M$ in the high fluence regime, as the increasing thermal load and demagnetization starts to counteract the helicity-dependent effect (grey shaded regions).

The most striking observation is the presence of strong helicity-dependent differences even for the off-resonant excitation at photon energies where the XUV spectroscopy shows very small or even zero XMCD (compare Fig. 1b for 51.0 and 64.0 eV). In general, the XMCD provides a potential source of helicity-dependent dynamics due to the interaction of circularly polarized XUV pulses and the angular momentum of the excited electrons leading to a polarization-dependence of the absorbed fluence. This dichroic absorption effect would manifest in a different amount of thermally induced demagnetization, depending on the helicity of the XUV excitation pulse. However, due to the lack of sizeable dichroic absorption in case of 51.0 and 64.0 eV (compare Fig. 1b), this mechanism can already be ruled out as the driving force behind the observed helicity-dependence at these two photon energies, which strongly indicates the existence of an opto-magnetic effect such as the IFE. This interpretation is further corroborated by the linear scaling with the fluence (XUV intensity) observed at 51.0 and 64.0 eV (white shaded regions in Fig. 3b), resembling a typical IFE feature, since the magnetization induced by the IFE depends linearly on the incident light intensity via the opto-magnetic, wavelength-dependent constant (see Eq. 1 and Battiato et al.[8]). A possible explanation for the more complex fluence-dependence and the deviation from a purely linear scaling observed at 54.1 and 56.1 eV can be given by the steep slope of the XAS and XMCD spectra (see Fig. 1b) at these photon energies, significantly changing the response of the system when the spectrum is slightly reshaped or shifted while the system is driven out of equilibrium, depending on the strength of the excitation[25,26].

The generation of helicity-dependent effects that are longer-lived compared to the duration of the driving pump pulse (see Fig. 2) have been previously reported in various experimental studies of the IFE on different materials[9,13,15,27–29]. Although the IFE is active only during the pump pulse duration, the subsequent magnetization evolution is affected on much longer time scales, especially, when the impulsively IFE-induced magnetization is large (here we reach a maximum of ≈30% of the ground-state $M_0$ values) and, consequently, the system needs longer times to accommodate such large magnetic changes. Particularly for high fluence excitation, the helicity-dependent effects therefore persist until the end of the pump-probe delay range (-2.2 to 30 ps) scanned in the experiment (for a full set of time traces as a function of fluence, see Supplementary Figure 4 and Supplementary Note 2).

*Ab initio* calculations of the inverse Faraday effect

To compare our experimental results to theoretical predictions of an opto-magnetic effect in the XUV spectral range, we carried out first *ab initio* calculations of the IFE[6,8] in vicinity of the Fe $M_{3,2}$ resonance. To reduce the computational complexity, the system was modeled as a stoichiometric $GdFe_2Co$ unit cell, focusing on the static opto-magnetic response for left ($\sigma_-$) and right ($\sigma_+$) circularly polarized excitation. The difference $\Delta_{IFE} = \mathcal{K}^{IFE}_{\sigma_-}(\omega) - \mathcal{K}^{IFE}_{\sigma_+}(\omega)$ between the calculated opto-magnetic constants is shown in Fig. 4 (right axis) in comparison to the largest helicity-dependent effects that were experimentally observed for the studied excitation photon energies (left axis, see also Fig. 3b). Note that this comparison is purely qualitative, as it is based on two non-equivalent quantities, namely the difference between the static opto-magnetic constants vs. the transient demagnetization amplitudes obtained at a certain time after excitation. A fully quantitative comparison cannot be made yet, as the constants $\mathcal{K}^{IFE}_{\sigma_\pm}(\omega)$ are obtained from stationary IFE calculations, so far not taking into account the fluence-dependent, non-equilibrium state present upon femtosecond excitation. To obtain a qualitative agreement with the experimental data, the calculated spectrum of the IFE response had to be shifted by +2.5 eV, which can be attributed to the effect of the core holes that is not included in the calculations[30–32] (see also the Methods section). The calculations support our experimental findings regarding the existence of an opto-magnetic effect at these wavelengths by predicting a highly wavelength-dependent IFE in vicinity of the $M_{3,2}$ resonance. The calculated $\Delta_{IFE}$ reverses its sign going through the resonance, which is again in qualitative agreement to the different polarity of $\Delta M$ observed in the experiment for below- and above-resonant excitation. This sign reversal occurs because the IFE spectra $\mathcal{K}^{IFE}_{\sigma_-}(\omega)$ and $\mathcal{K}^{IFE}_{\sigma_+}(\omega)$ display an energy shift with respect to each other at the 3p→3d resonance, generated by the exchange splitting of the 3d states. The very steep slope of the $\Delta_{IFE}$ spectrum around the sign reversal thereby coincides with the fluence-dependent sign change of the helicity-dependent effect $\Delta M$ observed for 56.1 eV excitation. As mentioned before, even a small but ultrafast spectral shift of the Fe $M_{3,2}$ resonance, which has been experimentally observed[25,26] for laser-driven non-equilibrium states in similar magnetic materials, can lead to a sign reversal of the IFE depending on the excitation fluence. Furthermore, it has to be emphasized that the calculations show a finite response at the photon energies 51.0 and 64.0 eV, corroborating the existence of a helicity-dependent interaction at photon energies where a thermal mechanism due to the XMCD effect could already be ruled out. However, the calculations are so far not able to quantitatively explain the experimental results regarding the magnitude of the observed helicity-dependence. Although the XAS show only a minor contribution due to the small amount of Co in the alloy (compare Fig. 1b at ≈62.0 eV), the IFE calculations show a finite extension of the opto-magnetic response up to the Co $M_{3,2}$ resonance, which can qualitatively explain the observation of helicity-dependent effects at 64.0 eV, but not yet the relatively large magnitude of the effect compared to the Fe response.

**Atomistic spin dynamics simulations**

In case of fully resonant excitation at 54.1 and 56.1 eV, the observed helicity-dependence could further result from the combined action of an IFE together with the dichroic absorption caused by the finite XMCD present at these photon energies (compare Fig. 1b). To quantify the actually expected effect of the XMCD on the helicity-dependent dynamics, atomistic spin dynamics (ASD) simulations were carried out[11,33–35], simulating the influence of the dichroic absorption on the helicity-dependent demagnetization amplitudes (see Fig. 4). For best comparability, the excitation fluence was calibrated between experiment and simulation by matching the demagnetization amplitudes obtained upon linearly polarized excitation. While the absence of any helicity-dependence in the XMCD-driven ASD simulations is obvious for photon energies without XMCD (51.0 and 64.0 eV), the simulation shows that even in the fully resonant case (54.1 and 56.1 eV), the expected dichroic demagnetization effects would be smaller by at least a factor of six (comparing $\Delta M_{sim}$ and $\Delta M_{exp}$ at 54.1 eV) and by almost an order of magnitude compared to the largest effect observed across the spectrum. Thus, a significant contribution of a purely thermal mechanism (via dichroic absorption) on the observed helicity-dependent dynamics can be ruled out, strengthening our conclusions regarding the existence of an opto-magnetic effect like the IFE for both on- and off-resonant excitation in vicinity of the Fe $M_{3,2}$ resonance.

**Conclusions**

In conclusion, we studied the impact of femtosecond XUV photoexcitation of a ferrimagnetic GdFeCo layer using circularly polarized FEL pulses at the Fe $M_{3,2}$ resonance. The systematic investigation of time-resolved magneto-optical Faraday rotation data obtained as a function of XUV excitation fluence, photon energy and polarization reveals a strong and long-lived helicity-dependent effect on the ultrafast demagnetization dynamics. The analysis of its wavelength- and fluence-dependence in comparison to static XMCD spectroscopy and time-resolved ASD simulations shows that the observed effect cannot be induced by the XMCD, i.e., thermally due to dichroic absorption. Instead, it has the characteristic fingerprints of an opto-magnetic effect, which is further corroborated by first *ab initio* theory on the IFE at the Fe $M_{3,2}$ resonance, showing qualitative agreement between the calculated IFE spectrum and the experimental data. Further experimental studies and theoretical efforts are needed for obtaining a quantitative understanding of the experimental results, especially with respect to the non-trivial fluence-dependence observed for excitation on the XUV resonance. This requires considering the IFE also in the highly non-equilibrium state upon femtosecond pulsed excitation, which is currently not included in the *ab initio* opto-magnetic theory. Complementary measurements upon resonant excitation of the Fe L edges, exciting 2p core-level electrons with even larger SOC compared to the 3p states, would provide further insights into the actual scaling of the IFE with the SOC and photon energy. Modern x-ray free-electron lasers can provide the required highly intense, circularly polarized femtosecond pulses of soft x-rays[20,36,37]. The observation of large opto-magnetic effects in the XUV spectral range also implies a significant, yet to be explored impact on ultrafast magnetization switching phenomena in addition to the MCD-induced helicity-dependence known from the visible-light regime[38] (regarding an IFE in the visible-light regime, see also Supplementary Note 4). Moreover, the different nature of the Gd 4f and Fe 3d magnetic moments with respect to their SOC and predominantly localized vs. itinerant character suggests that comparing element-specific excitation of the rare-earth and transition metal sublattices will contribute to fully revealing the microscopic mechanisms behind the XUV and soft x-ray IFE.

From a more general perspective, our findings reveal an efficient method to transiently generate large macroscopic magnetization on ultrafast time scales. Microscopically rooted in the stimulated Raman scattering process, the XUV-induced IFE demonstrated here can be seen, depending on the helicity of the exciting XUV pulse, to either generate or annihilate a transient magnetization within a range of up to ≈30% of the ground-state magnetization value of ferrimagnetic GdFeCo. Given its strength and robustness, such an opto-magnetic core-level effect is expected to be of relevance for the fields of ultrafast magnetism and spintronics, coherent magnetization control and nonlinear x-ray science. Moreover, touching upon the generality of the effect, the IFE via core-level excitations could essentially be extended to any nonmagnetic materials with strong spin-orbit coupling in order to transiently generate a macroscopic magnetic moment.

## Methods

### Experimental techniques and data acquisition

The time-resolved studies were performed at the DiProI end-station of the FERMI FEL[39], using a pump-probe geometry as schematically shown in Fig. 1a. For simultaneously measuring both the magneto-optical Faraday rotation in transmission and the Kerr rotation in reflection, two polarization-sensitive balanced photodetection schemes were employed. The magnetic contrast was obtained upon flipping an out-of-plane magnetic field of ±8 mT that was magnetically saturating the sample, also restoring the initial magnetization state after each pump-probe cycle. Taking the difference between the two traces recorded for opposite magnetization states leads to a quantity that is directly proportional to the magnetization, as the non-magnetic background cancels out. The experimental time resolution was evaluated to be ≈280 fs, using the method described by Ziolek et al.[40], accounting for the enclosed angle of 45° between pump and probe beams as well as their spatial footprint on the sample. The FEL and optical laser system were operating at 50 Hz repetition rate. The effective repetition rate of the XUV pump was reduced to 25 Hz by seeding only every second electron bunch in the FEL, allowing for an interleaved measurement to record both pumped and unpumped states of the sample. Supplementary Figure 1 exemplarily shows the data as recorded, i.e., the individually acquired magneto-optical Faraday signals as a function of pump-probe delay. Sorting the data after FEL pulse energy, which was shot-by-shot recorded by an $I_0$ gas monitor detector, allows retrieving the fluence-dependence (compare Supplementary Figure 2). For a more extensive technical background about the experimental methods and data acquisition, see Supplementary Notes 1 and 2.

**Sample design and spectroscopy**

The studied sample was a 20 nm thin amorphous film of a ferrimagnetic $Gd_{24}Fe_{67}Co_9$ alloy with an out-of-plane magnetic anisotropy. For static absorption spectroscopy, it was deposited on a 30 nm $Si_3N_4$ membrane, allowing transmission of the XUV radiation. For the time-resolved measurements, a 1 mm thick glass substrate was used in order to allow simultaneous detection of the Faraday and Kerr signals. In both cases, the magnetic sample was sandwiched between Ta seed and capping layers to protect it from oxidation. The exact sample compositions are Ta(2 nm)/GdFeCo(20 nm)/Ta(5 nm)/$Si_3N_4$(30 nm) and Ta(2 nm)/GdFeCo(20 nm)/Ta(5 nm)/Glass(1 mm), respectively.

The static XUV absorption and XMCD of the sample shown in Fig. 1b was characterized employing the ALICE reflectometer[41] at the PM3 beamline of BESSY II[42]. The XMCD spectrum thereby corresponds to the asymmetry ($A$) of the transmitted circularly polarized XUV radiation ($T_{\uparrow,\downarrow}$) measured for opposite magnetic field directions perpendicular to the sample plane, calculated by $A = (T_\uparrow - T_\downarrow) / (T_\uparrow + T_\downarrow)$. For obtaining the XAS ($B_{\uparrow,\downarrow}$), the transmitted XUV intensity was normalized to the beamline spectrum ($T_0$) by $B_{\uparrow,\downarrow} = -\ln(T_{\uparrow,\downarrow} / T_0)$. For comparison, this characterization was also carried out for a similar $Gd_{24}Fe_{76}$ sample, i.e., without the Co ingredient but otherwise identical composition. The comparison is shown in Supplementary Figure 8, showing the influence of the small fraction of Co atoms on the XAS and XMCD. To correct for the limited degree of circular polarization approaching the lower photon energy limits of the beamline, the magnetic part $\Delta\beta$ of the absorptive refractive index was calculated from the measured XMCD for the effective thickness of the Fe content within the alloy and normalized to reference XMCD measurements on pure Fe systems[43]. The corrected XMCD magnitude at the Fe $M_{3,2}$ resonance agrees with the values reported in the literature for comparable GdFe systems[44,45].

**Atomistic spin dynamics simulations**

Atomistic spin dynamics (ASD) simulations were carried out[11] to simulate the influence of the dichroic absorption on the helicity-dependent demagnetization amplitudes. The magnitude of the dichroic absorption was set according to the XMCD asymmetry obtained from the static XUV and XMCD spectroscopy (see Fig. 1b), i.e., (−6.2±0.5) % in case of 54.1 eV, (−3.0±0.5) % in case of 56.1 eV and (0.5±0.5) % in case of 51.0 eV excitation, respectively (note that for 64.0 eV, the XMCD is fully zero, thus no dichroic absorption is expected). The resulting simulated fluence-dependencies of the maximum demagnetization amplitudes upon $\sigma_\pm$-polarized excitation are shown in Supplementary Figure 9 together with their difference $\Delta M$.

Simulating also the dynamics excited by *linearly* polarized radiation allows better comparability between experiment and simulation by comparing the simulated XMCD-induced and experimentally obtained helicity-dependence $\Delta M$ based on the respective fluences that lead to the same amount of demagnetization in the linearly polarized case. Both the experimentally observed (yellow color) and simulated (purple color) $\Delta M$ values are shown in Supplementary Figure 10 as a function of the maximum demagnetization amplitudes induced by linearly polarized excitation of the same fluence. The comparison between experiment and ASD simulations presented in Fig. 4 is based on the largest $\Delta M$ observed in the experiment for each XUV photon energy and the interpolated value of $\Delta M$ taken from the ASD simulation resembling the respective magnitude of XMCD.

The ASD simulations are done using the VAMPIRE software package[33,34] based on the Landau-Lifshitz-Gilbert (LLG) equation. The system Hamiltonian is given by:

$$\mathcal{H} = -\sum_{i<j}(\mathbf{S}_i \mathbf{J}_{ij} \mathbf{S}_j) - k_\mathrm{u} \sum_i (\mathbf{S}_i \cdot \mathbf{e})^2, \qquad (2)$$

where $\mathbf{S}_i$, $\mathbf{S}_j$ are the normalized spin vectors on $i,j$ sites, $\mathbf{J}_{ij}$ is the exchange constant and $k_\mathrm{u}$ is the uniaxial magnetocrystalline anisotropy energy constant per site. We used similar parametrization as reported in literature[12,35]: $\mu_\mathrm{Fe} = 1.92\mu_\mathrm{B}$ and $\mu_\mathrm{Gd} = 7.63\mu_\mathrm{B}$, the anisotropy is taken as $k_\mathrm{u} = 8.07246 \cdot 10^{-24}$ J and the exchange interactions are $J_\mathrm{Fe\text{-}Fe} = 2.835 \cdot 10^{-21}$ J, $J_\mathrm{Gd\text{-}Gd} = 1.26 \cdot 10^{-21}$ J and $J_\mathrm{Fe\text{-}Gd} = -1.09 \cdot 10^{-21}$ J. We incorporate the rapid change in thermal energy of a system under the influence of a femtosecond laser pulse using the two-temperature model[46]:

$$T_e C_e \frac{dT_e(t)}{dt} = -G_{\text{e-ph}}\left[T_{\text{ph}}(t) - T_e(t)\right] + P_{\text{abs}}(t),$$

$$C_{\text{ph}} \frac{dT_{\text{ph}}(t)}{dt} = G_{\text{e-ph}}\left[T_e(t) - T_{\text{ph}}(t)\right],$$
(3)

where $C_e = 225$ Jm$^{-3}$K$^{-2}$, $C_{\text{ph}} = 3.1 \cdot 10^6$ Jm$^{-3}$K$^{-1}$ and $G_{\text{e-ph}} = 2.5 \cdot 10^{17}$ Wm$^{-3}$K$^{-1}$. The spin system is coupled to the electron temperature $T_e$ and the quantity $P_{\text{abs}}(t)$ corresponds to the laser power density absorbed by the electronic system and depends on laser fluence and light polarization. The XMCD effect is included by modifying $P_{\text{abs}}(t)$ proportional to the percentage obtained from the experiments.

### *Ab initio* calculations of the inverse Faraday effect

The *ab initio* calculations of the IFE were carried out using the second-order response theory formalism as derived by Berritta et al.[6] and Battiato et al.[8]. The calculations were performed with the full-potential, all-electron code WIEN2k[47], with spin-orbit interaction included.

The system was modeled as a stoichiometric GdFe$_2$Co unit cell in the AuCu$_3$ structure, i.e., as GdFe$_3$, but with one Fe atom replaced by Co. The lattice parameters were first optimized using the VASP code[48], after which the electronic structure and IFE was computed with the WIEN2k program. To capture the strong electron correlations in the Gd 4f shell, we employed the GGA+$U$ method, in the atomic limit version, with parameters $U = 7$ eV and $J = 1$ eV, in combination with the generalized gradient approximation (GGA)[49] to the exchange-correlation potential. The calculated band structure is shown in Supplementary Figure 11. The occupied Gd 4f states are located at $-6$ eV and the unoccupied 4f states start at $+4$ eV. The spin-polarized Fe and Co 3d states that are accessed by XUV photon excitation of the 3p states are in the energy window of 0 to 4 eV.

To compute the IFE, the exchange splitting of the 3p semi-core states of Fe and Co atoms was included as well[50]. The computed opto-magnetic IFE response is shown in Supplementary Figure 12. The photon energy- and helicity-dependent IFE constants are given for the total optically induced magnetization $M_{\text{ind}}$, i.e., they contain both spin and orbital contributions (see Berritta et al.[6]). The IFE constants for $\sigma_-$ and $\sigma_+$ circular polarizations are clearly distinct in the photon energy range of the Fe and Co M absorption edges while for energies outside of this range the distinction between $\sigma_-$ and $\sigma_+$ polarizations vanishes. The bottom panel of Supplementary Figure 12 shows $\Delta_{\text{IFE}}$, i.e., the difference between the opto-magnetic constants for the two helicities. To understand this spectrum, we can first note that, in the absence of spin magnetization in the alloy, the IFE spectra for σ− and σ+ polarizations would be identical but with opposite sign. This feature can be approximately recognized in Supplementary Figure 12. When there is spin magnetization present, the exchange splitting of the 3d valence bands leads to an energy-dependent shift in the individual IFE spectra, such that $\mathcal{K}_{\sigma_+}^{\text{IFE}}(\omega) \neq -\mathcal{K}_{\sigma_-}^{\text{IFE}}(\omega)$. A pronounced difference between $\mathcal{K}_{\sigma_+}^{\text{IFE}}(\omega)$ and $-\mathcal{K}_{\sigma_-}^{\text{IFE}}(\omega)$ occurs at 53−55 eV (top panel of Supplementary Figure 12), i.e., where the 3p→3d resonance occurs for Fe in the calculations. Around 58−60 eV, a similar difference is expected to appear near the Co 3p absorption edge. However, there will be an overlap of the M edges of Fe and Co in this photon energy range which can partially cancel out each other. In addition, we mention that previously, it was found that the *ab initio* computed XUV spectra of 3d transition metals compared well with measured spectra, but the computed energy onset of the $M_{3,2}$ edge differs from the measured onset position by about 2.5 eV[30–32]. This difference occurs because the *ab initio* calculated energy positions of the semi-core levels can deviate from the real energy positions and, in addition, there is an effect of the core hole in the XUV excitation which can lead to a shift in the binding energy of the semi-core levels[30,50]. The $\Delta_{\text{IFE}}$ spectrum shown in Fig. 4 has therefore been shifted by +2.5 eV to align with the experimental onset of the Fe M edge, which precisely corresponds to the shift that has been used in previous work[32]. We remark further that the computed IFE spectrum is valid in the quasi-static approximation, i.e., assuming that the computed IFE response is modulated only by the envelope of the XUV pulse. The employed second-order response theory is furthermore expected to be valid for reasonable, but not too high laser fluences.

## Acknowledgements

I.R. acknowledges funding from the Federal Ministry of Education and Research (BMBF) through project 05K16BCA (Femto-THz-X) and the European Research Council through project TERAMAG (Grant No. 681917). C.v.K.S., P.M.O. and S.E. would like to thank the German Research Foundation (DFG) for funding through CRC/TRR 227 projects A02 and MF (project ID 328545488). S.R. gratefully acknowledges the support of ARCHER UK National Supercomputing Service via the project e733. S.R. and T.O. gratefully acknowledge the financial support from the EPSRC TERASWITCH project (project ID EP/T027916/1). L.S. and P.M.O. acknowledge support by the Swedish Research Council (VR), the

K. and A. Wallenberg Foundation (Grants No. 2022.0079 and 2023.0336), and the European Union's Horizon 2020 Research and Innovation Programme under FET-OPEN Grant Agreement No. 863155 (s-Nebula). Part of the calculations were provided by the Swedish National Infrastructure for Computing (SNIC), funded by VR through grant No. 2018-05973. The authors acknowledge Elettra Sincrotrone Trieste for providing access to its free-electron laser facilities and thank all members of the different departments at FERMI for their outstanding assistance during the preparation and realization of the experiment. The authors also thank the Helmholtz-Zentrum Berlin für Materialien und Energie for the allocation of synchrotron-radiation beamtime.

**Author contributions**

I.R. conceived the original experimental idea. M.H. and I.R. conceived the experiment. The time-resolved measurements were prepared and performed by M.H., I.R., C.v.K.S., K.Y., E.J., B.V., V.C., K.L., F.C., D.N., E.P. and I.L.Q.. I.P.N., L.R. and G.D.N. helped setting up the experiment. On-site data monitoring was performed by M.H., K.Y., K.L. and F.C.. D.E. prepared and optimized the samples, M.H. carried out the static sample characterization. The data treatment and analysis was performed by M.H., with contributions from I.R. and F.C.. L.S. and P.M.O developed the IFE theory and carried out the *ab initio* calculations. S.R., R.C. and T.O. performed the ASD simulations. C.v.K.S., B.P. and S.E. contributed to the discussion and interpretation of the results. The manuscript was written by M.H. and I.R. with contributions from S.R. and P.M.O.. All authors commented on the manuscript.

**Competing interests**

The authors declare no competing interests.

**Data Availability**

The raw data were generated at the FERMI free-electron laser (DiProI end-station) and BESSY II synchrotron radiation (PM3 beamline) facilities. All data needed to evaluate the conclusions of this study are presented in the main article and/or the Supplementary Information, and are available from the corresponding authors upon request.

**Figures**

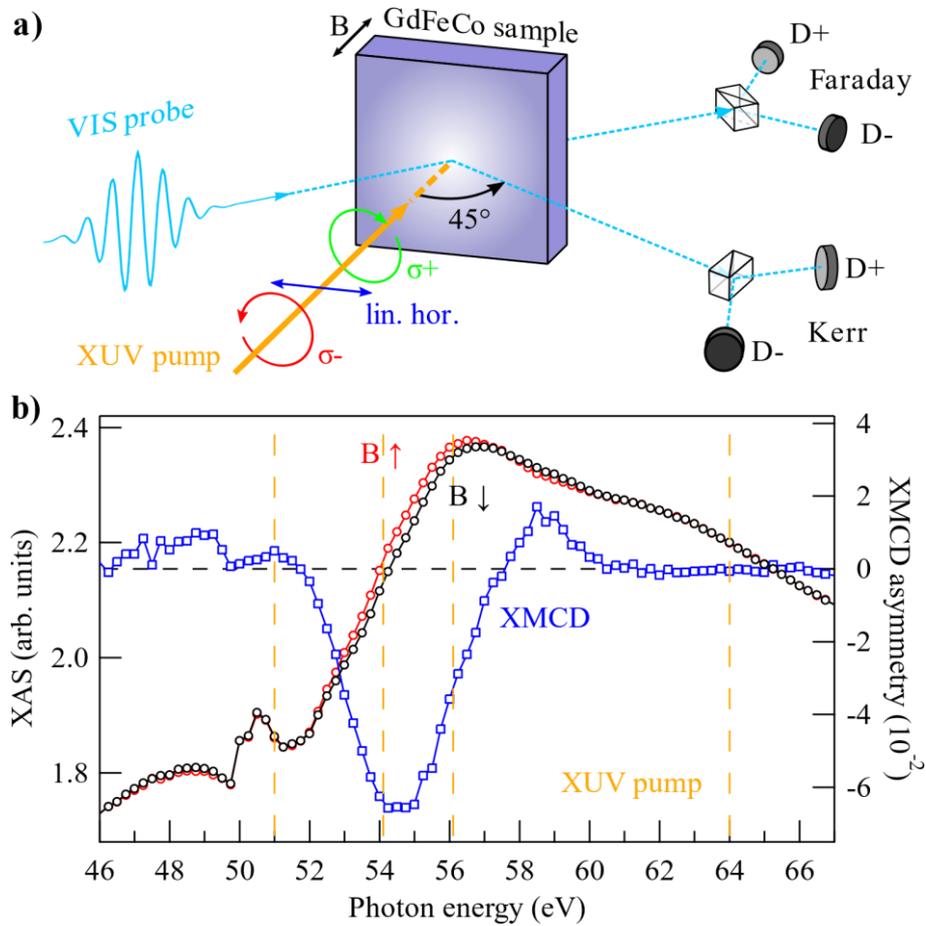

**Fig. 1. Experimental concept and spectroscopy.** (**a**) Schematic of the time-resolved pump-probe experiment implemented at FERMI. The GdFeCo sample is excited using extreme ultraviolet (XUV) pump pulses with variable polarization (circular polarization with opposite helicity $\sigma_\pm$ and linear horizontal). The visible (VIS) light pulses probe the magneto-optical Faraday and Kerr rotation using a balanced photodetection scheme, consisting of a Wollaston prism and two photodiodes ($D_\pm$). Magnetic contrast is obtained by flipping an out-of-plane magnetic field (B) applied to the sample. (**b**) Static XUV absorption (XAS, red and black circles for opposite magnetic fields, left scale) and x-ray magnetic circular dichroism (XMCD) asymmetry spectra (blue squares, right scale) of the GdFeCo sample measured across the Fe $M_{3,2}$ resonance. The yellow dashed lines indicate the photon energies used for XUV pumping (51.0, 54.1, 56.1 and 64.0 eV). The static spectroscopy was carried out with an energy resolution of ≤0.01 eV using the ALICE reflectometer[41] at the PM3 beamline of BESSY II[42].

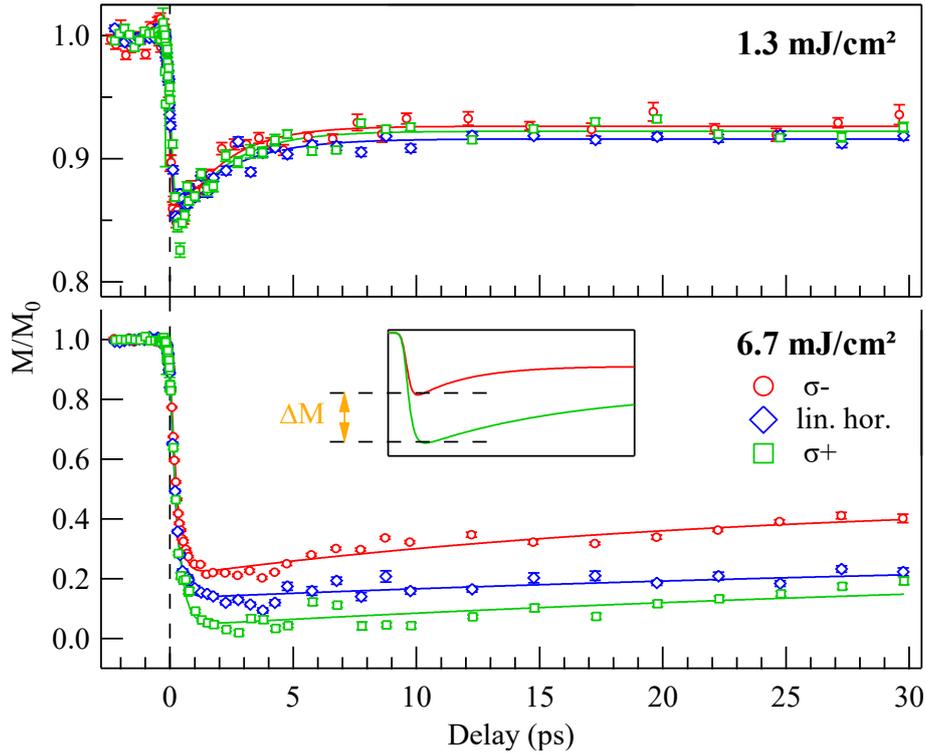

**Fig. 2. Time-resolved and helicity-dependent dynamics.** Magnetization dynamics induced by fs extreme ultraviolet (XUV) pulses with variable polarization (circular polarization with opposite helicities $\sigma_\pm$, red circles and green squares, and linear horizontal polarization, blue diamonds) for two different fluences 1.3 and 6.7 mJ/cm² at 54.1 eV photon energy, i.e., at the maximum x-ray magnetic circular dichroism (XMCD) effect − see Fig. 1b. The magnetization is normalized to the equilibrium magnetization in the unexcited state ($M/M_0$). The solid lines correspond to double-exponential fits − see also Supplementary Note 2, where the complete set of fit parameters is provided. The data sets measured for the full range of XUV fluences as well as the time-resolved differences between the magnetization transients obtained for $\sigma_\pm$-polarized excitation are shown in Supplementary Figures 3 and 4. The inset depicts how the helicity-dependent demagnetization amplitudes and their difference $\Delta M$ are obtained. The error bars are calculated as the standard error of the mean.

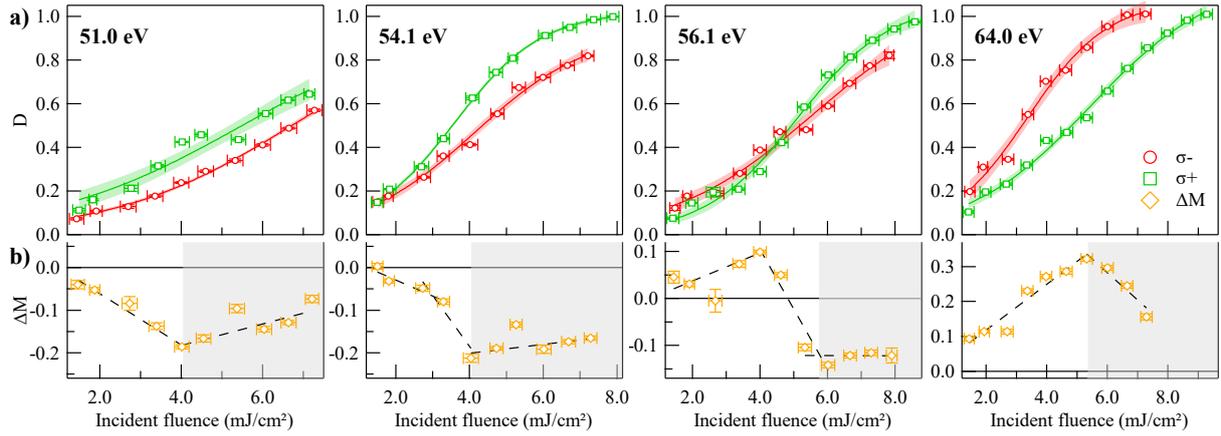

**Fig. 3. Fluence-dependence of the helicity-dependent dynamics.** (**a**) Demagnetization amplitudes (*D*) measured upon pumping with circularly polarized extreme ultraviolet (XUV) pulses with opposite helicities ($\sigma_\pm$, red circles and green squares) and (**b**) their difference $\Delta M$ (yellow diamonds) as a function of incident fluence and XUV photon energy. All values are normalized to the magnetization in the unexcited state and $\Delta M$ was obtained as shown in the inset of Fig. 2. The solid lines in panel **a** correspond to sigmoid functions fitted as guide to the eye; the shaded areas denote the experimental uncertainty by a 90% confidence interval. The dashed lines in panel **b** indicate linear trends fitted to the different fluence regimes, in which the magnitude of the helicity-dependence $\Delta M$ increases upon increasing the XUV fluence and subsequently saturates or even decreases for high fluences (white and grey shaded background). For clarity, only the fluence-dependence of the $\sigma_\pm$-polarized excitation is shown (for linear polarization, see Supplementary Figure 6). The error bars of the *D* values correspond to the uncertainty of the fitted demagnetization amplitudes (y-axis) and the standard error of the mean fluence values obtained by sorting the data after the free-electron laser (FEL) pulse energy (x-axis). The error bars of the $\Delta M$ values are calculated from the error propagation.

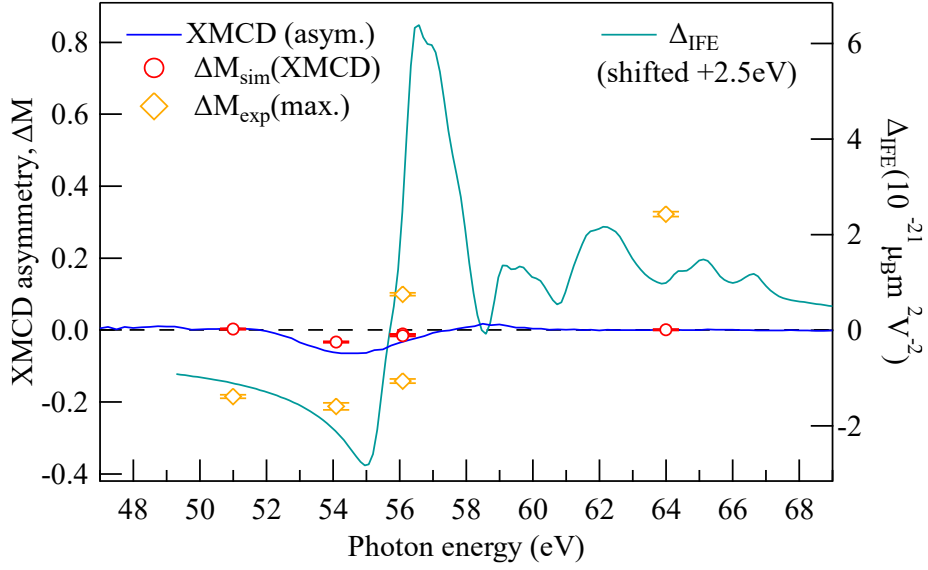

**Fig. 4. Comparison of experiment and theory.** Qualitative comparison of the experimentally observed difference in demagnetization amplitudes $\Delta M_{exp}$ (yellow diamonds, left scale) to the difference $\Delta_{IFE}$ (turquoise line, right scale) between the calculated opto-magnetic constants $\mathcal{K}^{IFE}_{\sigma_{\pm}}(\omega)$ obtained from *ab initio* inverse Faraday effect (IFE) theory. The red circles (left scale) correspond to the output of atomistic spin dynamics (ASD) simulations, simulating the expected demagnetization difference $\Delta M_{sim}$ induced by dichroic absorption due to the x-ray magnetic circular dichroism effect (XMCD, blue line). At 56.1 eV, two values are shown for $\Delta M_{exp}$, reflecting its opposite polarity in the low vs. high fluence regime, which coincides with the steep slope of the $\Delta_{IFE}$ spectrum at this photon energy. The error bars of $\Delta M_{sim}$ reflect the uncertainty of the static XMCD characterization and ASD simulations, corresponding to the standard deviation of different simulation runs for a $\pm 0.5\%$ variation of the input values of the XMCD magnitude. The error bars of $\Delta M_{exp}$ are calculated from the error propagation of the demagnetization amplitudes shown in Fig. 3.

SUPPLEMENTARY INFORMATION

# Ultrafast Opto-magnetic Effects in the Extreme Ultraviolet Spectral Range

Martin Hennecke,[*] Clemens von Korff Schmising, Kelvin Yao, Emmanuelle Jal, Boris Vodungbo, Valentin Chardonnet, Katherine Légaré, Flavio Capotondi, Denys Naumenko, Emanuele Pedersoli, Ignacio Lopez-Quintas, Ivaylo P. Nikolov, Lorenzo Raimondi, Giovanni De Ninno, Leandro Salemi, Sergiu Ruta, Roy Chantrell, Thomas Ostler, Bastian Pfau, Dieter Engel, Peter M. Oppeneer, Stefan Eisebitt, Ilie Radu[*]

[*]  hennecke@mbi-berlin.de; ilie.radu@xfel.eu

**Contents**

Supplementary Notes:

1. Experimental techniques and data acquisition
2. Data sorting and data treatment
3. Faraday vs. Kerr probing
4. Comparison to circularly polarized VIS/NIR-pumping

Supplementary Figures 1 to 13

Supplementary Table 1 to 5

**Supplementary Note 1: EXPERIMENTAL TECHNIQUES AND DATA ACQUISITION**

The time-resolved pump-probe measurements shown in the main article were carried out at the DiProI end-station using the FEL-1 beamline of the seeded free-electron laser (FEL) FERMI[1], employing the magneto-optical Faraday and Kerr effect to measure the time evolution of the GdFeCo sample magnetization as a function of XUV excitation photon energy, polarization and fluence. The magnetization dynamics of the sample were probed under an incidence angle of 45°, employing linearly polarized, ≈90 fs (FWHM), 400 nm optical pulses generated by a frequency-doubled Ti:sapphire laser system. Since the laser system shares an oscillator with a similar Ti:sapphire laser that provides the seeding pulses for initiating the FEL lasing process, the XUV pump and optical probe pulses are intrinsically synchronized and the jitter between the two pulses is reduced to ≈10 fs (see Ref. 2). Both Faraday and Kerr rotation of the polarization axis of the probing pulses were measured simultaneously in transmission and reflection geometry, respectively, by two independent polarization-sensitive detection setups using Wollaston prisms and balanced photo diodes. An electromagnet applying a saturating magnetic field of ±8 mT perpendicular to the sample plane was used to restore the initial magnetization state of the sample after each pump-probe cycle. Additionally, the magnetic field was flipped every 200 pulses to obtain a magnetic contrast corresponding to the difference between the Faraday and Kerr signals measured for opposite magnetic field directions. The delay between pump and probe pulses was adjusted via an optical delay stage in the probe beam path. Both the FEL and the optical laser system were running at a repetition rate of 50 Hz. By seeding only every second accelerated electron bunch in the modulator section of the FEL, the XUV pump repetition rate is effectively reduced to 25 Hz, allowing for an interleaved measurement of the pumped and unpumped states of the sample. Fast oscilloscopes triggered by the time base of the FEL were used to record and split the pumped and unpumped signals of the balanced photo diodes. The acquired Faraday rotation signals, i.e., the pumped and unpumped state for opposite magnetic field directions as a function of pump-probe delay and XUV polarization, are shown exemplarily in Supplementary Figure 1 for an incident excitation fluence of 4.7 mJ/cm² at a photon energy of 64.0 eV.

In order to excite the sample using different XUV photon energies, the wavelength of the FEL was adjusted by either changing the FEL seeding laser wavelength and the undulators gap between the magnetic sections, or by changing the harmonic order of the emitted radiation. Supplementary Table 1 shows the FEL parameters used in the experiment. A spectrometer in the XUV beam path was used to record the spectrum of the FEL shots in order to determine the spectral bandwidth, i.e., the energy resolution, by fitting the spectrum with a Gaussian function. The FEL pulse durations were determined according to Ref. 3, scaling with the seeding laser pulse length of ≈170 fs and inversely with the harmonic order of the FEL. The resulting pulse lengths show only negligible dependence on the XUV wavelengths used in our experiment and are thus approximated by ≈90 fs. At each photon energy, the polarization of the XUV was alternated between linear horizontal, $\sigma_-$ and $\sigma_+$ by moving the undulator of the FEL. The degree of circular polarization after transmission through the DiProI beamline was characterized in Ref. 4 and shown to be consistently above 90% for both $\sigma_-$ and $\sigma_+$ up to XUV wavelengths of 60 nm, approaching almost ≈100% in the lower wavelength range below 25 nm. Thus, tuning the XUV wavelength in a range from 19.37 to 24.31 nm, any change in helicity due to the different wavelengths is expected to be on the order of ≈1% or less, thus not significantly impacting the polarization state of the XUV pulses. Previous pump-probe studies reported by the authors of Ref. 5, which were carried out at the same end station utilizing a similar wavelength and fluence range, have further demonstrated the very high reproducibility and intensity correlation when switching the helicity of the FEL radiation between $\sigma_-$ and $\sigma_+$. Changing the incident pump fluence on the sample was accomplished by attenuating the FEL pulse energies using solid-state aluminum filters of different thicknesses (100−500 nm) for rough adjustments and a variable pressure inside of a gas absorber for fine-tuning. The resulting attenuation, average pulse energy and shot-to-shot fluctuations of the FEL could be monitored by two $I_0$ gas monitor detectors (GMD) before and after the attenuator.

For optimizing the pump-probe conditions, the spot sizes (FWHM) of the FEL pump and optical probe pulses were adjusted and measured directly in the sample plane by covering parts of the sample with a layer of fluorescent paint and profiling the beam spots on a camera, leading to an uncertainty of ≈5−10% in FWHM due to this method. The XUV pump spot size was adjusted to 300×300 µm² using a Kirkpatrick-Baez (KB) mirror focusing system in front of the experimental chamber. The spot size of the probing laser was tuned to 85×85 µm² via optical lenses. The delay between pump and probe pulses was adjusted via an optical delay stage in the probe beam path. For spatial overlap, the probing spot was centered within the much larger pump spot, in order to probe a homogeneously pumped area. Additionally, a YAG screen placed before the KB mirrors was used to validate the XUV beam size and position before the focusing optics. Spot sizes and pump-probe overlap were checked regularly throughout the experiment in order to assure reliable and stable pump-probe conditions and minimize any systematic error that could emerge from temporal drifts or a change of FEL parameters, especially after changing photon energy and polarization.

## Supplementary Note 2: DATA SORTING AND DATA TREATMENT

The recorded time-resolved Faraday and Kerr data were sorted by incident excitation fluence using the pulse energies of the FEL shots recorded by the $I_0$-GMD. Shot-to-shot fluctuations of the FEL source lead to a statistical distribution of pulse energies around an average value targeted by the attenuator settings. Thus, each data set collected during a single pump-probe delay scan contains a large amount of fluence-dependent information that can be extracted.

Supplementary Figure 2 shows exemplary the histograms of FEL pulse energies per shot of three consecutive pump-probe delay scans recorded at different average excitation fluences. Sorting the data by pulse energy using a step size of 1 µJ and averaging only over the Faraday and Kerr probe belonging to the same interval of (x±0.5) µJ allows obtaining a fluence-dependence with a high density of points. The binning window widths, resulting in an uncertainty of the excitation fluence, were chosen as a trade-off between the density of points in the fluence diagram and the signal-to-noise ratio of the averaged Faraday and Kerr signals which is limited by the number of shots that fall into the intervals. Taking also the beamline transmission of ≈60% into account, the pulse energies can be divided by the XUV spot size to obtain the incident fluence in units of mJ/cm² on the sample. Due to the shape of the statistical pulse energy distribution, the data points are generally not equally distributed within each binning window, especially at the rising and falling edges of the distribution (compare the histograms in Supplementary Figure 2). The average fluence of the data points within a binning window is thus not necessarily located at its center. As the fluctuations are random and the shape of the fluence distribution slightly varied between subsequently recorded scans for different polarization and photon energies, the pairs of averaged data points for σ₋ and σ₊ excitation that were subtracted for the determination of Δ*M* (as shown in Fig. 3 of the main article) possess a slight fluence deviation. The corresponding increased uncertainty of Δ*M* on the fluence axis is reflected by the error bars, taking into account the fluence mismatch as well as the initial fluence uncertainty via error propagation.

Supplementary Figure 3 shows the full data set of the polarization- and fluence-dependent magnetization dynamics induced by resonant 54.1 eV excitation as a function of pump-probe delay, obtained from sorting the time-resolved magneto-optical Faraday rotation after XUV pump pulse energy. Supplementary Figure 4 shows the corresponding difference between the magnetization transients obtained for σ₋ and σ₊ excitation, i.e., the helicity-dependent effect, as a function of pump-probe delay and excitation fluence. The data shows that especially in the high fluence regime (≥4 mJ/cm²), where the IFE-induced effects are large, the helicity-dependent effect persists until the end of the pump-probe delay range (-2.2 to 30 ps) scanned in the experiment. As discussed in the main article, such long-lived effects on the magnetization are reasonable, as the system needs much longer times compared to the pump pulse duration in order to accommodate the large IFE-induced magnetization changes (up to 30% of equilibrium magnetization). Also for excitation fluences ≥4 mJ/cm² and within the experimental uncertainty, the magnitude of the helicity difference either slowly increases further after the initial sub-picosecond rise or stays constant within this time interval (see, e.g., the data for 4 mJ/cm² and 6 mJ/cm²). This behavior, taking place on a ten picoseconds time scale, can qualitatively be attributed to different recovery times depending on the demagnetization amplitudes after σ₋ or σ₊ excitation, i.e., the magnetization level from which the system has to recover back to the initial state. However, this does not affect the results or conclusions presented in the main article, where the helicity-dependent effect is quantified based on Δ$M$ values that are obtained from the maximum demagnetization amplitudes at early times (≤2 ps), as indicated in the inset of Fig. 2 of the main article.

The data shown in Supplementary Figures 3 and 4 was fitted using a double exponential fit function (solid lines), taking into account the ultrafast demagnetization (Fig. 3) or the helicity-dependent magnetization changes (Fig. 4), the subsequent relaxation process and the experimental time resolution:

$$f(t) = g(t) \otimes \begin{cases} A, & t \leq 0 \\ A - B\left[1 - \exp\left(-\frac{t}{\tau_B}\right)\right] + C\left[1 - \exp\left(-\frac{t}{\tau_C}\right)\right], & t > 0 \end{cases} \quad (1)$$

with $A$ corresponding to the unpumped Faraday signal at negative delays, $B$ and $C$ to the amplitudes of the two exponential components modeling the ultrafast rise or drop and subsequent relaxation processes, and $\tau_B$ and $\tau_C$ to the respective exponential time constants. By convolution with a Gaussian function $g(t)$, the experimental time resolution of ≈280 fs (FWHM) is taken into account, which was determined using the method described in Ref. 6 from the pulse duration of the pump and probe pulses as well as the experimental geometry, i.e., the angle between the two beams and their footprints on the sample.

The obtained exponential time constants of the ultrafast demagnetization process are shown in Supplementary Figure 5a as a function of excitation fluence and XUV polarization. The corresponding time constants of the rising edge of the helicity-dependent effect are shown in Supplementary Figure 5b. The complete set of normalized fit parameters (i.e., with A=1 and A=0, respectively, corresponding to the unpumped state) are provided in Supplementary Tables 2-5. Please note that in the low fluence regime, the limited temporal resolution of the experiment (≈280 fs) in conjunction with the low amplitudes of demagnetization and helicity-dependent effect leads to a large uncertainty. It also has to be noted that, apart from the low fluence regime, the scanned pump-probe delay interval does not contain sufficient amount of the recovery dynamics for an accurate fit, which leads to a large uncertainty in the determination of the relaxation amplitudes and time constants.

## Supplementary Note 3: FARADAY VS. KERR PROBING

Supplementary Figures 6 and 7 show the maximum demagnetization amplitudes $D$ upon $\sigma_\pm$ and linearly polarized excitation ($D = 1 - \min[M/M_0]$) as well as the differences $\Delta M = D[\sigma_-] - D[\sigma_+]$ as a function of incident excitation fluence for the XUV excitation photon energies of 51.0, 54.1, 56.1 and 64.0 eV, comparing magneto-optical Faraday and Kerr probing. Note that all values are normalized to the equilibrium magnetization in the unexcited state. The data were fitted using sigmoid functions serving as a guide to the eye. The shaded areas in Supplementary Figures 6a and 7a correspond to a 90% confidence interval as an estimation of the experimental uncertainty. The shaded areas in Supplementary Figures 6b and 7b correspond to the different fluence regimes that within the experimental uncertainty indicate the helicity-dependent effect to scale almost linearly with the fluence (white area) until a saturated state is reached where it almost stays constant or starts to decrease again (gray area). The dashed lines serve as a guide to the eye and correspond to linear trends fitted to the different fluence regimes. Comparing the Faraday and Kerr data reveals a slight difference in demagnetization amplitudes, which can be related to the different information depths of the Faraday and Kerr measurements, corresponding either to the whole depth of the sample in transmission or the penetration depth of the optical light in reflection geometry, respectively. Furthermore, the helicity-dependent effect probed by the Kerr rotation undergoes a stronger attenuation in the saturated regime, as the measured demagnetization amplitudes approach the fully demagnetized state for lower excitation fluences compared to the Faraday data.

**Supplementary Note 4: COMPARISON TO CIRCULARLY POLARIZED VIS/NIR-PUMPING**

In order to *directly* compare the XUV-IFE to the IFE induced by VIS/NIR-pumping of the valence band, additional time-resolved measurements were performed on the very same GdFeCo sample that was used in the XUV-studies presented in the main article. Employing a table-top time-resolved MOKE setup (1.55 eV pump − 3.1 eV probe), the laser-induced demagnetization dynamics were systematically studied as a function of excitation fluence and helicity. The pump pulses were circularly polarized using a λ/4-wave plate, alternating their helicity between $\sigma_-$ and $\sigma_+$ with ≤0.1% fluence deviation. The results are shown in Supplementary Figure 13, revealing an helicity-dependent effect on the order of ≈1% of the equilibrium magnetization. The magnitude of the effect was obtained in the same way as for the XUV-IFE studies presented in the main article, i.e., from the difference $\Delta M$ between the maximum demagnetization amplitudes $D$ induced by the opposite helicities $\sigma_\pm$. These measurements show that the helicity-dependent effects induced by VIS/NIR light are by an order of magnitude smaller compared to the XUV-IFE measured on the same GdFeCo sample, strongly suggesting that the IFE scales indeed with the spin-orbit coupling, which is much stronger for the core-levels than for valence band electrons (e.g., 1.1 eV vs. 65 meV in case of Fe 3p and 3d electrons, respectively[7]).

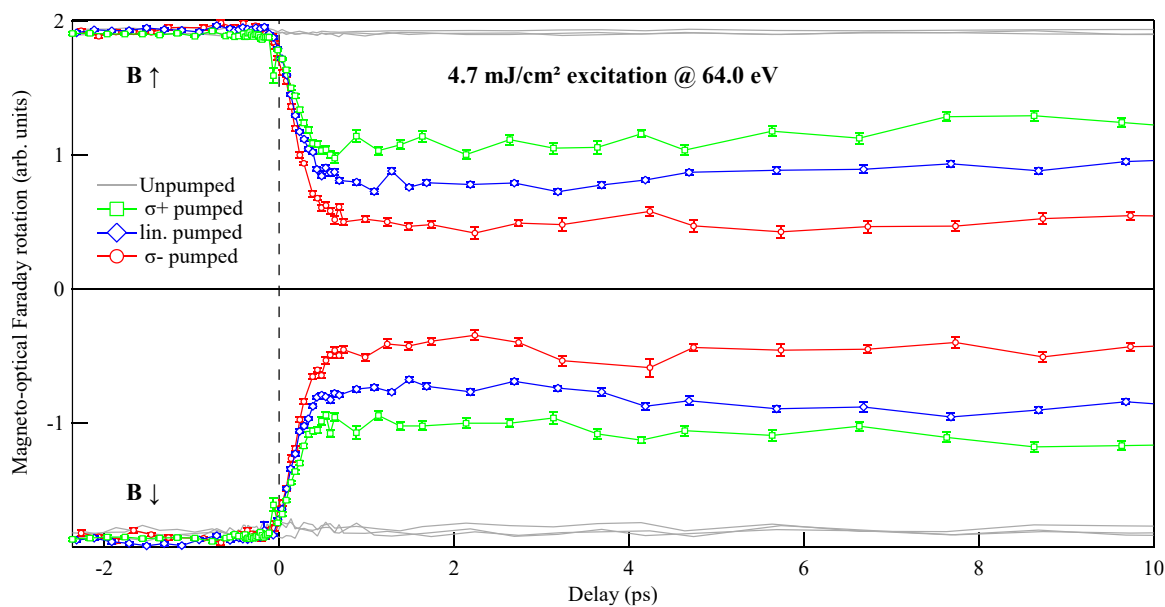

**Supplementary Figure 1.**
Transient magneto-optical Faraday rotation upon excitation with an incident fluence of 4.7 mJ/cm² and a photon energy of 64.0 eV, recorded for opposite magnetic field (B) directions as a function of pump-probe delay and XUV polarization. The pumped and unpumped states are acquired by probing at 50 Hz repetition rate while pumping at only 25 Hz. The error bars are calculated as the standard error of the mean.

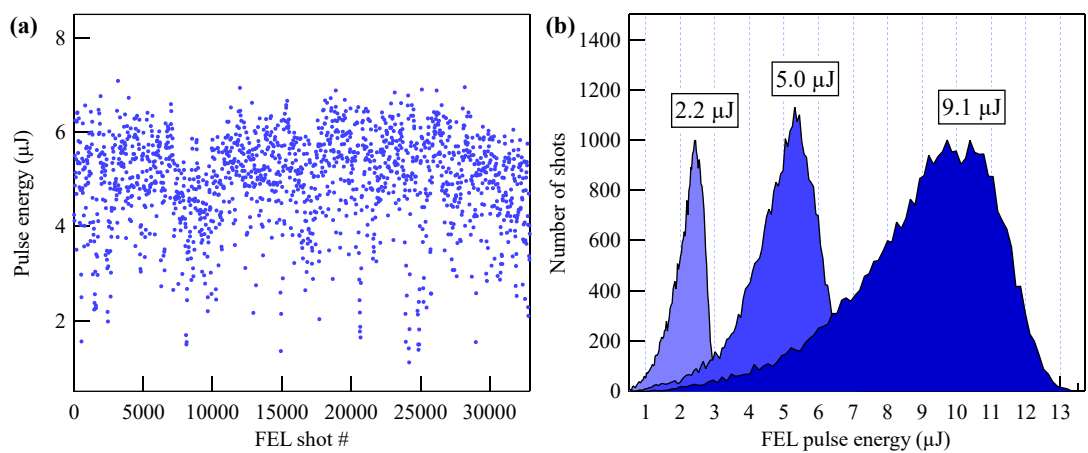

**Supplementary Figure 2.**
Analysis of the shot-resolved FEL pulse energies recorded by the $I_0$ gas monitor detector. The graph shows exemplary the data recorded for linearly polarized XUV radiation at a photon energy of 54.1 eV. (**a**) FEL pulse energies per shot as recorded for an average pulse energy of 5.0 μJ. Only every twentieth shot is plotted for better visibility. (**b**) Histograms of the statistical distribution of XUV pulse energies recorded during three subsequent pump-probe delay scans using different average excitation fluences. The Faraday/Kerr data was sorted by averaging only over those data points where the sample was excited by FEL shots with the same pulse energy, as defined by a 1 μJ grid (shown as dashed lines).

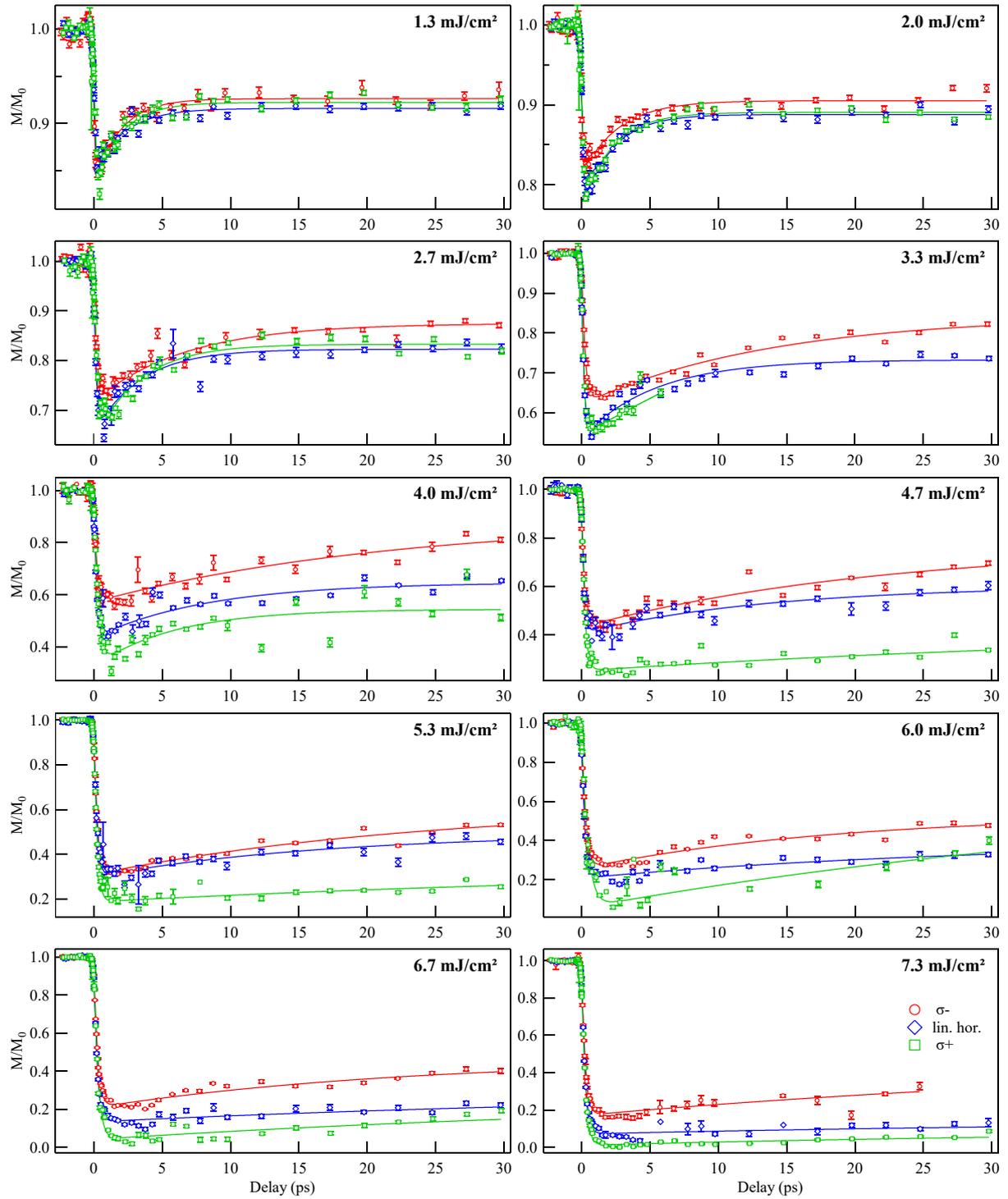

**Supplementary Figure 3.**
Transient magnetization dynamics of the system induced by XUV pulses of 54.1 eV photon energy, probed by the normalized magneto-optical Faraday rotation as a function of pump-probe delay, XUV polarization and incident excitation fluence. The magnetization is normalized to the equilibrium magnetization in the unexcited state ($M/M_0$). The error bars are calculated as the standard error of the mean.

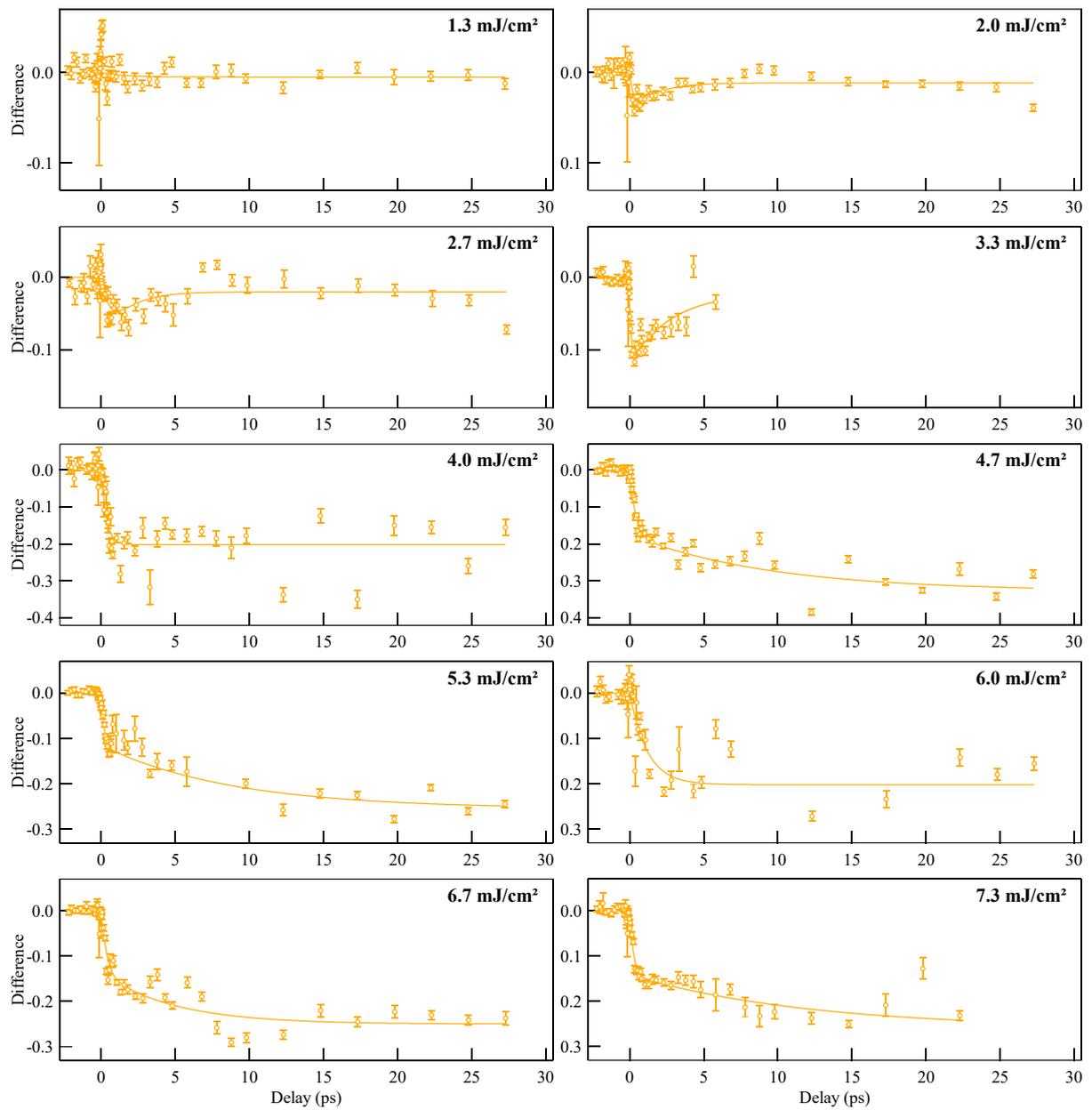

**Supplementary Figure 4.**
Difference between the magnetization transients obtained for σ− and σ+ excitation (data from Supplementary Figure 3), showing the helicity-dependent effect as a function of pump-probe delay and excitation fluence. The error bars are calculated from the error propagation of the standard error of the mean.

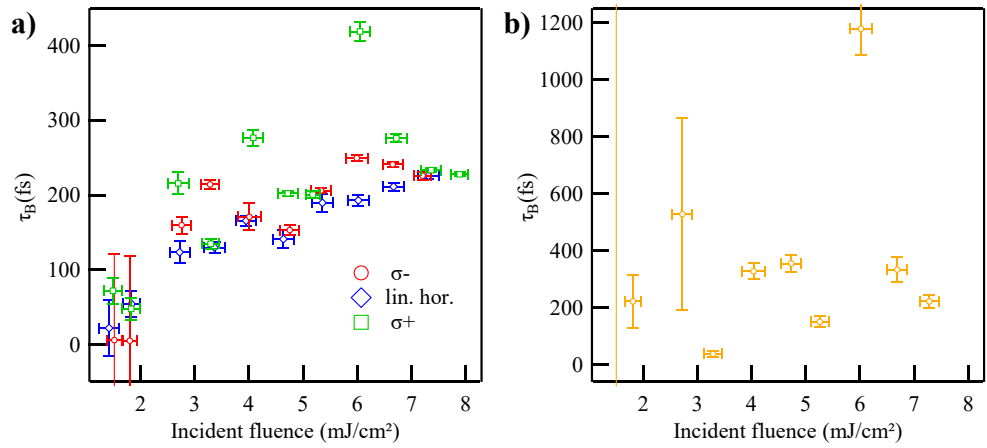

**Supplementary Figure 5.**
Fitted exponential time constants of the data presented in Supplementary Figures 3 and 4 as a function of incident excitation fluence. (**a**) Time constants of the ultrafast demagnetization induced by the three different XUV polarizations. (**b**) Time constants of the rising edge of the helicity-dependent effect. The error bars correspond to the uncertainty of the fitted exponential time constants (y-axis) and the standard error of the mean fluence values obtained by sorting the data after free-electron laser pulse energy (x-axis).

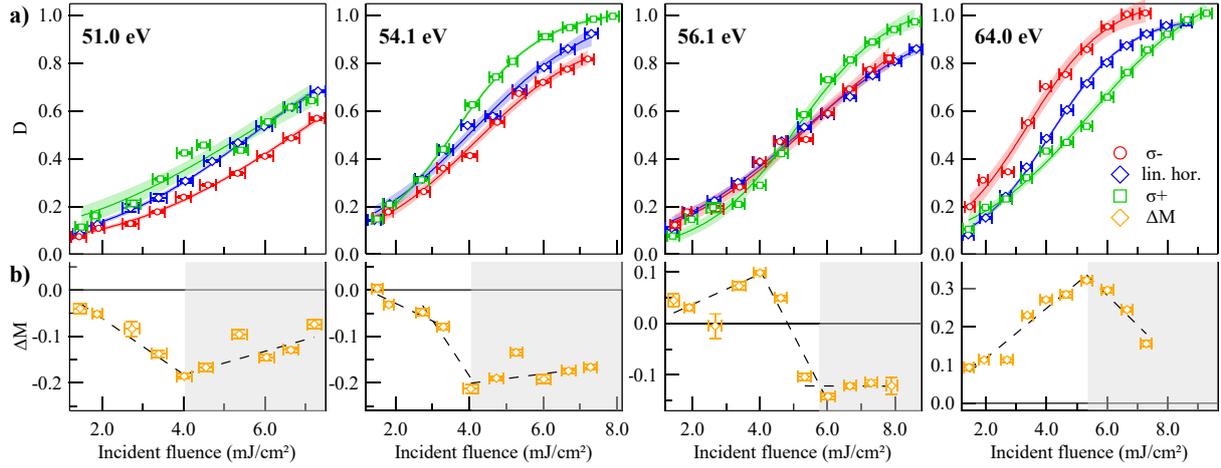

**Supplementary Figure 6.**
(**a**) Maximum demagnetization amplitudes (*D*) obtained from the transient magneto-optical *Faraday* rotation upon $\sigma_\pm$- and linearly polarized excitation (red circles, blue diamonds and green squares), and (**b**) their difference $\Delta M$ (yellow diamonds) as a function of incident fluence and XUV photon energy. The error bars of the *D* values correspond to the uncertainty of the fitted demagnetization amplitudes (y-axis) and the standard error of the mean fluence values obtained by sorting the data after the free-electron laser (FEL) pulse energy (x-axis). The error bars of the $\Delta M$ values are calculated from the error propagation.

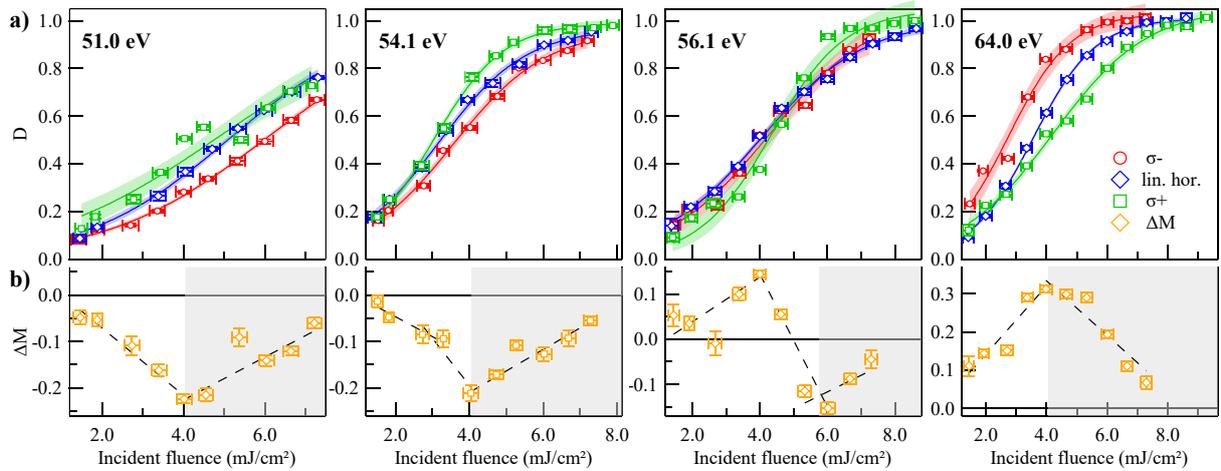

**Supplementary Figure 7.**
(**a**) Maximum demagnetization amplitudes (*D*) obtained from the transient magneto-optical *Kerr* rotation upon $\sigma_\pm$- and linearly polarized excitation (red circles, blue diamonds and green squares), and (**b**) their difference $\Delta M$ (yellow diamonds) as a function of incident fluence and XUV photon energy. The error bars of the *D* values correspond to the uncertainty of the fitted demagnetization amplitudes (y-axis) and the standard error of the mean fluence values obtained by sorting the data after the free-electron laser (FEL) pulse energy (x-axis). The error bars of the $\Delta M$ values are calculated from the error propagation.

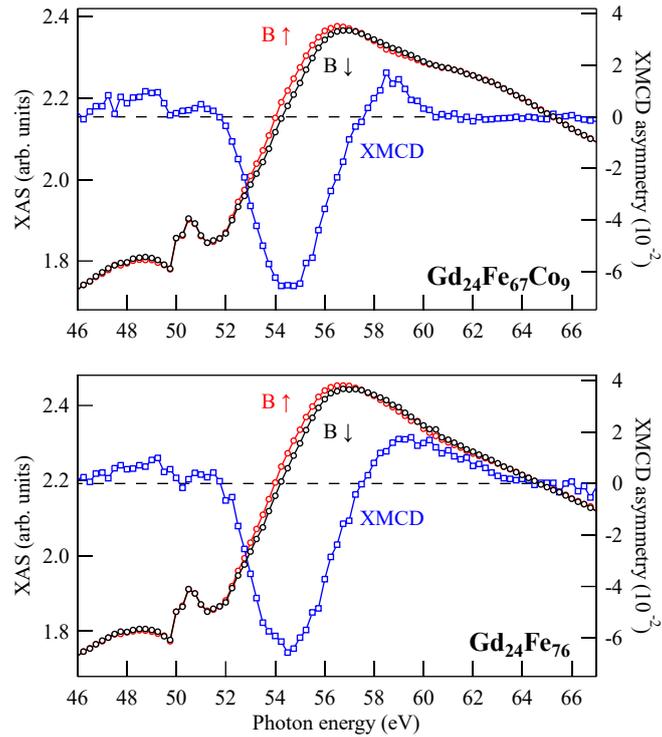

**Supplementary Figure 8.**
Static XUV absorption (XAS, red and black circles, left scale) and XMCD asymmetry spectra (blue squares, right scale) of the $Gd_{24}Fe_{67}Co_9$ and $Gd_{24}Fe_{76}$ samples, measured in the photon energy range of the Fe $M_{3,2}$ resonance. The small peak at 62.0 eV in the XAS of the GdFeCo sample arises due to resonant excitation of the small Co constituent, which is not present in the XAS of the GdFe sample. On the contrary, the XMCD of the GdFeCo sample is suppressed at 62.0 eV compared to GdFe, which can be attributed to the opposite polarity of the Fe and Co XMCD at this photon energy (see, e.g., Ref. 8 for XMCD spectra of elemental Fe and Co), cancelling out each other in their superposition due to the different concentration of Fe and Co in the alloy.

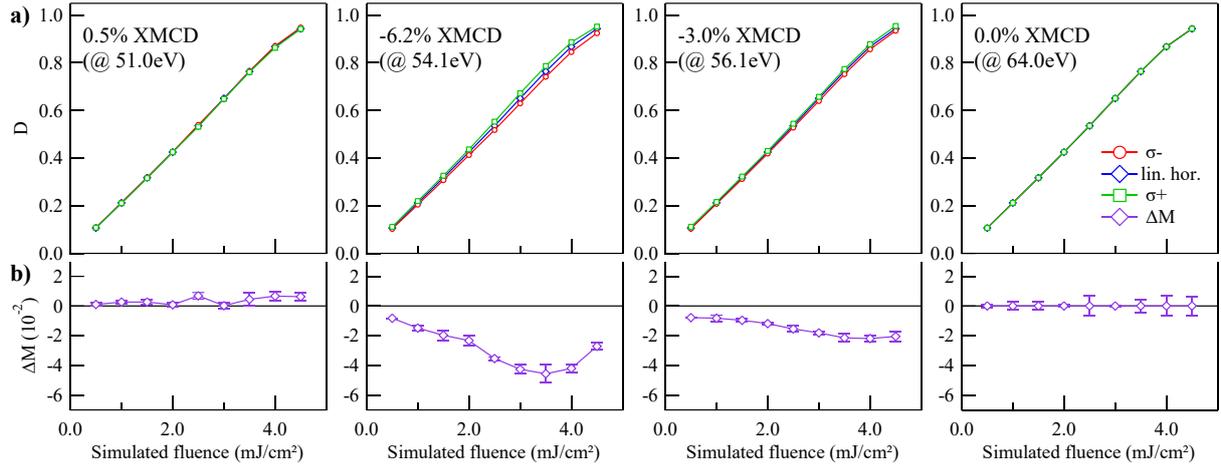

**Supplementary Figure 9.**
(**a**) Maximum demagnetization amplitudes (*D*) obtained from ASD simulations for $\sigma_\pm$- and linearly polarized excitation (red circles, blue diamonds and green squares) as a function of simulated excitation fluence and magnitude of XMCD, that was set according to the static XMCD spectroscopy of the sample, and (**b**) the corresponding difference $\Delta M$ (purple diamonds). The error bars are obtained by varying the input values for the XMCD magnitude by $\pm 0.5\%$.

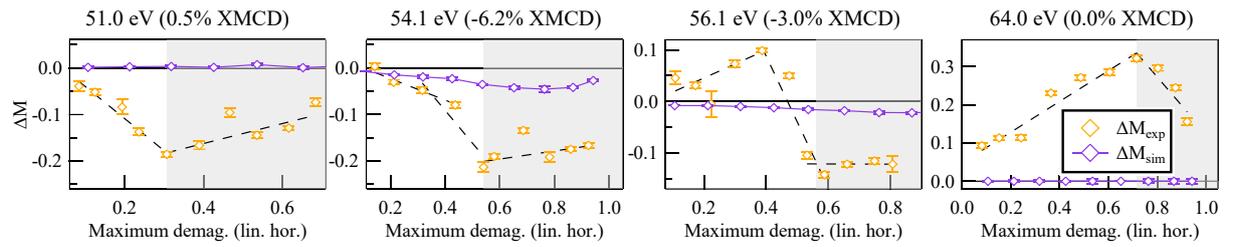

**Supplementary Figure 10.**
Experimentally observed (yellow color, compare Supplementary Figure 6) and simulated (purple color, compare Supplementary Figure 9) $\Delta M$ values as a function of the maximum demagnetization amplitudes induced by linearly polarized excitation of the same fluence. The error bars of $\Delta M_\text{sim}$ are obtained by varying the input values for the XMCD magnitude by $\pm 0.5\%$. The error bars of $\Delta M_\text{exp}$ are calculated from the error propagation of the demagnetization amplitudes shown in Supplementary Figure 6.

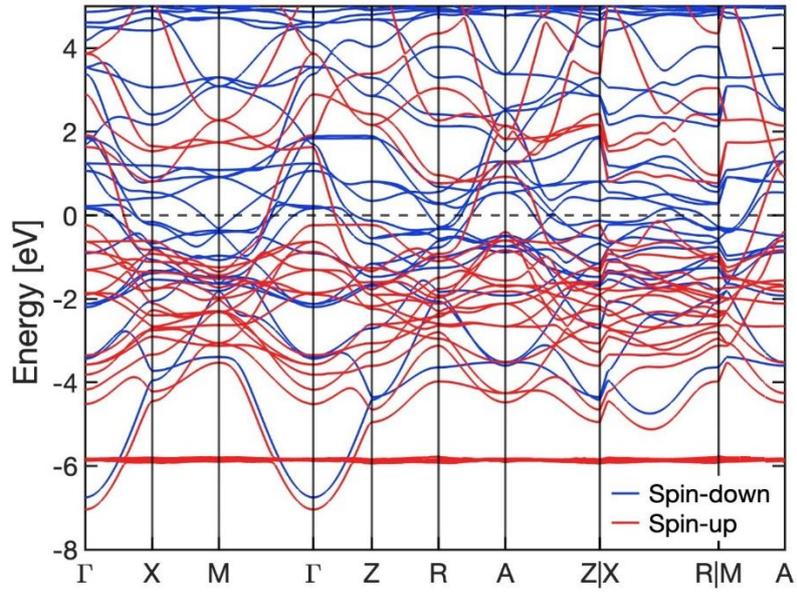

**Supplementary Figure 11.**
Calculated spin-polarized band structure of GdFe$_2$Co along high-symmetry lines of the simple tetragonal Brillouin zone, with special symmetry points as indicated.

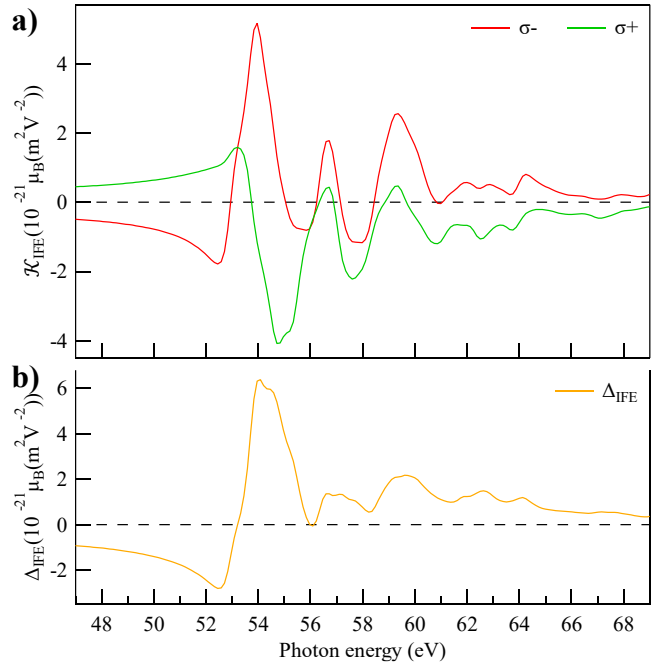

**Supplementary Figure 12.**
Static *ab initio* calculations of the Inverse Faraday Effect (IFE) constant $\mathcal{K}^{\text{IFE}}$ in the XUV spectral range for a GdFe$_2$Co unit cell. (**a**) The calculated total opto-magnetic response to σ$_\pm$-polarized XUV radiation as a function of photon energy, i.e., the sum over the spin (*S*) and orbital (*L*) responses to left (σ$_-$) and right (σ$_+$) circularly polarized excitation, respectively. (**b**) The corresponding difference $\Delta_{\text{IFE}}$ between the total opto-magnetic constants, which is also shown in Fig. 4 of the main article for qualitative comparison to the experimental data.

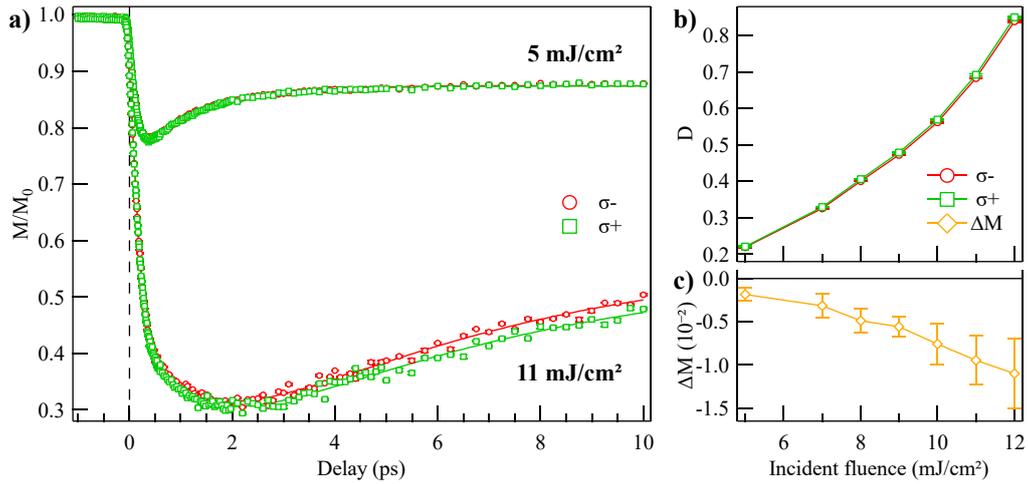

**Supplementary Figure 13.**
(**a**) Laser-induced magnetization dynamics ($M/M_0$) driven by circularly polarized fs NIR-pulses (1.55 eV) with opposite helicities ($\sigma_\pm$) for two different fluences 5 and 11 mJ/cm². The error bars are calculated as the standard error of the mean. (**b**) Maximum demagnetization amplitudes (*D*) measured upon $\sigma_\pm$-pumping and (**c**) their difference $\Delta M$ as a function of incident fluence. All values are normalized to the equilibrium magnetization in the unexcited state. The error bars of the *D* values correspond to the uncertainty of the fitted demagnetization amplitudes. The error bars of the $\Delta M$ values are calculated from the error propagation.

| Energy (eV)  | Wavelength (nm) | FEL seed (nm)      | Pulse length (fs) |
|--------------|-----------------|--------------------|-------------------|
| 51.00±0.02   | 24.31±0.01      | 243.14 (10. harm.) | ≈92               |
| 54.10±0.02   | 22.92±0.01      | 251.90 (11. harm)  | ≈89               |
| 56.10±0.02   | 22.10±0.01      | 243.14 (11. harm.) | ≈89               |
| 64.00±0.04   | 19.37±0.01      | 251.90 (13. harm.) | ≈84               |

**Supplementary Table 1.**
XUV photon energies and wavelengths used for excitation by tuning the FEL seeding laser wavelength and harmonics. The spectral bandwidth (energy resolution) is given by fitting the XUV spectrometer measurements with a Gaussian peak and taking the FWHM of the peak. The pulse lengths are approximated from the seeding laser pulse duration of ≈170 fs and its harmonic order[3].

| Fluence (mJ/cm²) | B | $\tau_B$ (ps) | C | $\tau_C$ (ps) |
|---|---|---|---|---|
| 1.3 | 0.165±0.006 | 0.006±0.116 | -0.091±0.006 | 1.92±0.15 |
| 2.0 | 0.188±0.004 | 0.005±0.113 | -0.094±0.004 | 2.67±0.15 |
| 2.7 | 0.283±0.004 | 0.160±0.011 | -0.157±0.004 | 6.67±0.46 |
| 3.3 | 0.379±0.002 | 0.214±0.007 | -0.223±0.005 | 13.8±0.8 |
| 4.0 | 0.428±0.006 | 0.171±0.018 | -0.316±0.030 | 22.0±4.3 |
| 4.7 | 0.569±0.004 | 0.153±0.006 | -0.339±0.025 | 21.9±3.0 |
| 5.3 | 0.691±0.002 | 0.206±0.003 | -0.310±0.018 | 24.4±2.6 |
| 6.0 | 0.744±0.003 | 0.249±0.004 | -0.280±0.014 | 18.6±2.0 |
| 6.7 | 0.793±0.003 | 0.241±0.003 | -0.260±0.022 | 22.7±3.3 |
| 7.3 | 0.829±0.005 | 0.226±0.006 | -0.328±0.323 | 50.0±61.2 |

**Supplementary Table 2.**
Fitted parameters of the transient magnetization dynamics upon σ₋-polarized excitation.

| Fluence (mJ/cm²) | B | $\tau_B$ (ps) | C | $\tau_C$ (ps) |
|---|---|---|---|---|
| 1.3 | 0.151±0.003 | 0.022±0.038 | -0.067±0.003 | 2.25±0.18 |
| 2.0 | 0.231±0.005 | 0.054±0.017 | -0.120±0.004 | 2.21±0.13 |
| 2.7 | 0.347±0.007 | 0.123±0.015 | -0.170±0.007 | 3.35±0.24 |
| 3.3 | 0.463±0.004 | 0.130±0.007 | -0.196±0.004 | 5.31±0.28 |
| 4.0 | 0.565±0.004 | 0.166±0.007 | -0.212±0.005 | 8.35±0.61 |
| 4.7 | 0.593±0.008 | 0.141±0.012 | -0.196±0.016 | 13.9±3.0 |
| 5.3 | 0.702±0.007 | 0.190±0.012 | -0.196±0.022 | 16.7±4.3 |
| 6.0 | 0.792±0.005 | 0.193±0.007 | -0.195±0.063 | 30.2±15.1 |
| 6.7 | 0.866±0.004 | 0.211±0.006 | -0.175±0.179 | 50.0±67.2 |
| 7.3 | 0.928±0.004 | 0.226±0.005 | -0.084±0.217 | 50.0±169.0 |

**Supplementary Table 3.**
Fitted parameters of the transient magnetization dynamics upon lin. hor. - polarized excitation.

| Fluence (mJ/cm²) | B | $\tau_B$ (ps) | C | $\tau_C$ (ps) |
|---|---|---|---|---|
| 1.3 | 0.166±0.004 | 0.072±0.018 | -0.088±0.004 | 1.98±0.17 |
| 2.0 | 0.228±0.003 | 0.048±0.015 | -0.119±0.003 | 2.25±0.09 |
| 2.7 | 0.363±0.007 | 0.215±0.015 | -0.196±0.006 | 3.32±0.17 |
| 3.3 | 0.457±0.006 | 0.135±0.007 | -0.975±4.653 | 50.0±253.0 |
| 4.0 | 0.676±0.008 | 0.276±0.011 | -0.220±0.008 | 5.85±0.54 |
| 4.7 | 0.749±0.002 | 0.202±0.003 | -0.191±0.077 | 50.0±26.5 |
| 5.3 | 0.813±0.005 | 0.201±0.005 | -0.165±0.133 | 50.0±54.9 |
| 6.0 | 0.945±0.010 | 0.419±0.013 | -0.639±0.295 | 50.0±31.5 |
| 6.7 | 0.959±0.004 | 0.276±0.005 | -0.237±0.145 | 50.0±40.2 |
| 7.3 | 0.988±0.002 | 0.233±0.002 | -0.096±0.066 | 50.0±44.9 |

**Supplementary Table 4.**
Fitted parameters of the transient magnetization dynamics upon $\sigma_+$-polarized excitation.

| Fluence (mJ/cm²) | B | $\tau_B$ (ps) | C | $\tau_C$ (ps) |
|---|---|---|---|---|
| 1.3 | 0.005±0.002 | 1.50±1.98 | 0 | - |
| 2.0 | 0.044±0.012 | 0.222±0.091 | -0.033±0.012 | 1.64±0.66 |
| 2.7 | 0.097±0.103 | 0.529±0.338 | -0.077±0.102 | 1.57±1.23 |
| 3.3 | 0.119±0.005 | 0.038±0.011 | -0.098±0.016 | 2.82±1.08 |
| 4.0 | 0.202±0.004 | 0.328±0.028 | 0 | - |
| 4.7 | 0.174±0.008 | 0.355±0.029 | 0.153±0.008 | 9.37±1.86 |
| 5.3 | 0.115±0.005 | 0.150±0.020 | 0.140±0.007 | 8.78±1.67 |
| 6.0 | 0.202±0.005 | 1.18±0.09 | 0 | - |
| 6.7 | 0.136±0.009 | 0.334±0.043 | 0.114±0.009 | 4.84±0.94 |
| 7.3 | 0.141±0.006 | 0.222±0.025 | 0.119±0.023 | 11.9±5.4 |

**Supplementary Table 5.**
Fitted parameters of the difference between the transient magnetization dynamics upon $\sigma_-$- and $\sigma_+$-polarized excitation, i.e., of the helicity-dependent effect.

## Supplementary References